\documentclass{article}

\usepackage{PRIMEarxiv}

\usepackage[utf8]{inputenc}
\usepackage{fullpage}
\usepackage[dvipsnames, table]{xcolor}
\usepackage{amsfonts} 
\usepackage{amsmath}
\usepackage{dsfont}
\usepackage{bbm}
\usepackage{graphicx}
\usepackage{moreverb,url}
\usepackage{tikz}
\usepackage{natbib}
\usepackage{setspace}
\usepackage{comment}
\usepackage{multirow}
\usepackage{array}
\usepackage{booktabs}
\usepackage{hyperref}

\hypersetup{
    colorlinks,
    citecolor={blue}
}

\title{Multivariable Behavioral Change Modeling of Epidemics in the Presence of Undetected Infections}

\author{Caitlin Ward\textsuperscript{1,\thanks{Corresponding author: ward-c@umn.edu}}, Rob Deardon\textsuperscript{2, 3}, Alexandra M. Schmidt\textsuperscript{4}}

\begin{document}

\maketitle

\spacing{1.2}

\noindent \textsuperscript{1} Division of Biostatistics and Health Data Science, University of Minnesota \\
\textsuperscript{2} Faculty of Veterinary Medicine, University of Calgary \\
\textsuperscript{3} Department of Mathematics and Statistics, University of Calgary \\
\textsuperscript{4} Department of Epidemiology, Biostatistics, and Occupational Health, McGill University

\vspace{10mm}

\onehalfspacing

\begin{abstract}

Epidemic models are invaluable tools to understand and implement strategies to control the spread of infectious diseases, as well as to inform public health policies and resource allocation. However, current modeling approaches have limitations that reduce their practical utility, such as the exclusion of human behavioral change in response to the epidemic or ignoring the presence of undetected infectious individuals in the population. These limitations became particularly evident during the COVID-19 pandemic, underscoring the need for more accurate and informative models. To address these challenges, we develop a novel Bayesian epidemic modeling framework to better capture the complexities of disease spread by incorporating behavioral responses and undetected infections. In particular, our framework makes three contributions: 1) leveraging additional data on hospitalizations and deaths in modeling the disease dynamics, 2) accounting for data uncertainty arising from the large presence of asymptomatic and undetected infections, and 3) allowing the population behavioral change to be dynamically influenced by multiple data sources (cases and deaths). We thoroughly investigate the properties of the proposed model via simulation, and illustrate its utility on COVID-19 data from Montr\'{e}al and Miami.

\end{abstract}

\keywords{Bayesian inference \and Compartmental model \and Transmission model \and SIHRD \and SIR}

\section{Introduction}
Epidemic models are essential tools for understanding transmission dynamics and informing public health responses. Infectious disease transmission is influenced by many factors, including the infectiousness of the disease, the susceptibility of the population, and the behavior of the population in response to the outbreak. Behavioral change could include city- or state-wide lockdown policies, school closures, or personal changes such as voluntary masking and can result in a delayed or reduced peak or multiple waves. Despite the fact that human behavior drives disease transmission  \citep{arthur2017contact}, many traditional modeling approaches treat transmission parameters as fixed over time. The COVID-19 pandemic highlighted the limitations of this assumption, and motivated the development of models that explicitly incorporate behavioral feedback mechanisms. The incorporation of behavioral change in epidemic models can help public health officials better understand disease spread and the impact of interventions or messaging, ultimately leading to better outbreak control  \citep{osi2025simultaneous}.

Compartmental epidemic models describe infectious disease dynamics by dividing the population into epidemiologically important disease states. The foundational compartmental model of Kermack and McKendrick \citep{kermack1927} is the \textbf{S}usceptible, \textbf{I}nfectious, or \textbf{R}emoved (SIR) model, which specifies that individuals start in the susceptible compartment, become infected through contact with infectious individuals, and become removed when they recover with immunity or die from the disease. Parameters describe the flow between compartments, with the transition from susceptible to infectious being of primary interest as this captures disease transmission. These models can be specified deterministically through a series of ordinary differential equations, or incorporate stochasticity by probabilistically describing the parameters and/or the transitions between compartments. Stochastic models are more realistic representations of the spread of an infectious pathogen, and incorporating this uncertainty is particularly useful when describing disease spread at the start of an epidemic or in small populations  \citep{roberts2015nine}. Bayesian methods are commonly used to implement stochastic models where the primary goal is making inference on transmission parameters using real observed epidemic data. The Bayesian framework is advantageous in this setting due to its ability to provide uncertainty quantification that naturally accounts for parameter estimation, incorporate prior knowledge on some parameters, and impute missing data through data augmentation  \citep{o1999bayesian, lekone}. 


Several behavioral change SIR models have been proposed prior to the occurrence of the COVID-19 pandemic (see \cite{funk2010modelling} and \cite{verelst2016behavioural} for reviews of behavioral change modeling approaches published through 2015). The onset of the pandemic brought increased attention towards epidemic modeling, however, a recent review found only 38\% of models attempted to incorporate human behavior  \citep{lee2025review}. Additionally a large majority of the behavioral change literature both pre- and post-COVID-19 focused on model-specification and simulation, and not on performing parameter estimation using observed data  \citep{funk2010modelling, verelst2016behavioural, lee2025review}. In many cases, this is due to highly specified behavioral dynamics that are not supported by currently available data and therefore the associated behavioral parameters are difficult to estimate. More broadly, recent work has highlighted that many epidemic models incorporate behavior exogenously (e.g., through mobility or other external data sources), but this can lead to poor estimation of disease dynamics  \citep{rahmandad2022enhancing}. Thus, incorporating behavioral responses endogenously within an epidemic system has been identified as an important direction for improving estimation  \citep{hamilton2024incorporating}.

Recently a stochastic Bayesian behavioral change modeling framework was developed where transmission is dynamically modified by a so-called `alarm' function, which captures time-varying changes in transmission that may be due to the population engaging in protective behaviors  \citep{ward2023bayesian, ward2025framework}. The alarm is specified as a function of a single endogenous input (incidence or prevalence) and is inversely associated with transmission, with larger values of the alarm corresponding to greater engagement in protective behaviors and reduced transmission. With an endogenous input, this creates a feedback loop where increasing case counts leads to higher alarm and a subsequent reduction in transmission, but when case counts drop, the alarm function decreases leading to increased future transmission. Implementing these models using Bayesian methods allows for inference on transmission and behavioral parameters using observed data, facilitating new knowledge about transmission dynamics and the role of behavior in informing transmission during a previous outbreak. It has also been shown to provide superior model fit to outbreaks of COVID-19, Ebola, and foot and mouth disease, compared to models ignoring behavioral change. Despite this success, these existing models rely on simplifying assumptions: a behavioral alarm that depends on only one endogenous measure of disease severity and the use of a standard SIR structure.  

These simplifying assumptions may be reasonable in settings where data are limited or when the primary goal is short-term prediction. However, they may be restrictive in contexts such as COVID-19, where multiple high-quality data streams (e.g., cases, hospitalizations, and deaths) are available and where public risk perception may be shaped by multiple signals. The well-accepted Health Belief Model suggests that an individual's perceived risk depends jointly on perceived susceptibility to infection and perceived severity of disease outcomes \citep{champion2008health}. In the context of epidemic data, reported case counts provide a natural proxy for perceived susceptibility, while reported deaths capture perceived severity. Both have been shown to influence risk perception of COVID-19  \citep{lohiniva2022covid, gutierrez2022information}. Agent-based models of SARS, H1N1, and Ebola have incorporated perceived susceptibility based on recent case counts and severity based on recent deaths or the number of news reports  \citep{durham2012incorporating, durham2012deriving, yan2018impact}. However these models assumed that the difference components of risk perception independently contributed to behavioral change. This motivates extending the alarm-based behavioral change modeling framework to incorporate both cases and deaths. This also allows for the model to be formally parametrized to infer the relative contributions of perceived susceptibility and severity on behavioral change during an outbreak.

Extending the alarm to include deaths introduces an additional consideration on the compartmental structure of the model. As deaths are not directly generated in an SIR model, an alarm based on cases and deaths in that model would incorporate both endogenous and exogenous inputs. To maintain the endogenous formulation of the alarm used in previous work, we instead adopt a more detailed SIHRD compartmental structure that explicitly represents susceptible,infectious, hospitalized, recovered and deceased states. As hospitalization and death data are readily available for COVID-19, these disease states can more easily be incorporated into the stochastic Bayesian models without a huge increase in the computational complexity that sometimes limits these models  \citep{andersson2012stochastic}. Using only endogenous inputs preserves the feedback loop between transmission and behavior, improving the interpretability of the alarm process. Additionally, we hypothesize that this will lead to increased accuracy in estimation of important parameters, such as the reproductive number compared to a simplified compartmental structure with a mix of endogenous and exogenous alarm inputs.

Throughout the COVID-19 pandemic, modeling has been complicated by the large presence of undetected infections, either due to lack of testing availability, asymptomatic infection, or false negative test results  \citep{menachemi2020population, irons2021estimating}. Existing population-level models have addressed this by adding an undetected compartment  \citep{melis2021undetected, bhaduri2022extending}, using a multiplier on the observed case counts  \citep{hao2020reconstruction, ma2024epidemiological}, or using a capture-recapture framework to estimate a lower bound on the number of undetected infections  \citep{bohning2020estimating}. However, these approaches to incorporate undetected infections have not been considered in the context of endogenous behavioral change modeling, and may be important to include when behavioral responses are driven by the epidemic trajectory. We seek to develop a behavioral change framework that incorporates undetected infections to understand whether ignoring this missing data mechanism leads to bias in estimating either the transmission or behavioral parameters of interest.

Motivated by these challenges, we extend the stochastic Bayesian behavioral change model in three ways. First, we replace the SIR structure with an SIHRD model that explicitly represents hospitalizations and deaths. Second, we incorporate undetected infections. Third, we introduce a multivariable alarm that depends on both reported cases and deaths, allowing us to quantify their relative contributions to population behavioral change and enabling new scientific insights on how populations respond to different signals of epidemic risk and how these responses might change over the course of an epidemic. Our goal is to evaluate how these extensions meaningfully improve inference on both transmission and behavioral dynamics through simulation studies and application to COVID-19 data.

The remainder of the paper is organized as follows. In Section \ref{methods}, we review the univariable alarm SIR model and present our behavioral change SIHRD framework with a multivariable alarm and incorporating undetected infections. In Section \ref{simStudy}, the proposed methodology is rigorously investigated through simulations. An analysis of two waves of COVID-19 in Miami, Florida and Montr\'{e}al, Qu\'{e}bec is performed in Section \ref{analysis} to illustrate the utility and limitations of the proposed modeling approach on observed data. Concluding remarks are provided in Section \ref{disc}.

\section{Methods}\label{methods}

\subsection{Behavioral Change SIR Model} \label{methods_sir}

We start by briefly introducing the discrete time behavioral change SIR model of \cite{ward2023bayesian} that serves as the foundation for our work. Assume a closed population of size $N$ and let $S_t$, $I_t$, and $R_t$ denote the number of individuals in the susceptible, infectious, and removed compartments in the continuous time interval $[t, t+1)$, respectively. Transition vectors $I^*_{t}$ and $R^*_{t}$ represent the number of individuals entering the indicated compartment in this interval. Transitions are temporally described by
\begin{align*} 
    S_{t+1} &= S_t - I^*_t \\ 
    I_{t+1} &= I_t + I^*_t - {R_t}^*  \\ 
    R_{t+1} &= R_t +  {R_t}^*. 
\end{align*} 
Given counts in each compartment at time $0$ and the transition vectors, $\boldsymbol{S}$, $\boldsymbol{I}$, and $\boldsymbol{R}$  are fully determined by these difference equations. The initial states can be fixed or estimated using a multinomial prior. Transitions between compartments are binomially distributed as $I^*_{t} \sim Bin\left(S_{t}, \pi_t^{(SI)} \right)$ and $R^*_{t} \sim Bin\left(I_{t}, \pi^{(IR)} \right)$. 

Of primary interest is the transmission probability $\pi_t^{(SI)}$, traditionally specified as $\pi_t^{(SI)} = 1 - \exp \left(-\beta \frac{I_t}{N}\right)$. This probability is derived from assumptions of an independent Poisson contact process and constant infection probability given a contact, where $\beta$ captures both the contact rate and infection probability as these are not separately identifiable  \citep{brownReproductive}. Behavioral change is incorporated by allowing the transmission rate, $\beta$, to be modified by a time-varying level of alarm, $a_t$, in the population as
\begin{equation} \label{eq_pSI}
\pi_t^{(SI)} = 1 - \exp\left\{-\beta(1 - a_t) \frac{I_t}{N}\right\}.
\end{equation}
We chose the power alarm as it avoids identifiability issues that could arise from more complex alarm specifications. More details on this are provided in Section 6 of the Supplementary Material.

The alarm is constrained to be in the interval $[0, 1]$, and is interpreted as the proportional reduction in transmission due to the alarm in the population. When $a_t = 0$, the population is in its natural `unalarmed' state, and transmission is described only by $\beta$. When $a_t = 1$, the population is in its maximal alarmed state and transmission is reduced to zero. In \cite{ward2023bayesian}, the alarm was specified as a function of previously observed incidence smoothed over the past $m$ days, i.e., $a_t = f\left(I^*_{t-1, m} \right)$, where $I^*_{t-1, m} = \frac{1}{m} \sum_{i = t-m-1}^{t-1}I^*_i$ with smoothing parameter $m \in \{1, 2, ..., t-1\}$. Several functions have been considered to describe the alarm, and we will use the one parameter `power' alarm with growth rate $k>0$
\begin{equation} \label{eq_unialarm}
    a_t = 1 - (1 - I^*_{t-1, m}/N)^{1/k}.
\end{equation}

The removal probability is typically specified as $\pi^{(IR)} = 1 - \exp \left(-\gamma\right)$, which derives from an exponentially distributed length of the infectious period. However, this may be a restrictive assumption and without complete data on both $\boldsymbol{I^*}$ and $\boldsymbol{R^*}$ would require data-augmented MCMC approaches to implement  \citep{lekone}. To incorporate more flexible modeling of the infectious period and avoid data augmentation, we use the infectious duration-dependent (IDD) specification of  \citep{ward2022idd}, which was shown to be computationally efficient and to improve estimation relative to traditional data augmentation. Let $I_{wt}$ be the number of individuals on day $w$ of the infectious period at time $t$ and $T_I$ denote the fixed length of the infectious period. The IDD transmission probability is 
\begin{equation} \label{eq_pSI_IDD}
    \pi_t^{(SI)} = 1 - \exp \left\{-\beta(1 - a_t) \frac{\sum_{w = 1}^{T_I} f(w) I_{wt}}{N}\right\},
\end{equation}
where $f(w)$ is a function defining the weighted contribution of infectious individuals depending on how long they have been infectious. In our implementation, we use a logistic decay function with a maximum curve value of 1, $f(w) = 1/[1 + \exp\{\nu(w - w_0)\}]$, where $w_0$ provides the inflection point of the curve, and $\nu > 0$ provides the decay rate. As the length of the infectious period is fixed, removal times are fully specified as $R^*_{t + T_I} = I^*_t$, for all $t$. Therefore, the $R^*_{t}$ are no longer considered to be binomially distributed and we no longer need to estimate the removal probability $\pi^{(IR)}$ . Instead, the parameters of the logistic decay curve are used to quantify the length of time infectious individuals are transmitting the disease. The logistic decay curve assumes peak transmissibility at the start of the infectious period, which may not always be realistic. Alternative IDD curves can allow later peaks, but \cite{ward2022idd} found minimal changes in overall transmission estimation between various curves and the logistic decay curve was found to have the highest MCMC efficiency, defined as the number of effective samples of the reproductive number per unit time.

\subsection{Multivariable Alarm in an SIHRD model with Undetected Infections} \label{comp_models}

We extend the behavioral change SIR model in three important ways. First, we consider an expanded SIHRD compartmental structure to include hospitalizations (H) and deaths (D). The SIHRD model is temporally described by the following set of difference equations:
\begin{eqnarray*} 
    S_{t+1} &=& S_t - I^*_t \\ 
    I_{t+1} &=& I_t + I^*_t - H^*_t - {R_t^I}^*  \\ 
    H_{t+1} &=& H_t + H^*_t - {R_t^H}^* -D^*_t \\ 
    R_{t+1} &=& R_t +  {R_t^I}^* + {R_t^H}^*\\
    D_{t+1} &=& D_t + D^*_t,
\end{eqnarray*}
where $I_t^*$, $H_t^*$, and $D_t^*$ correspond with the observed incidence, new hospitalizations, and new deaths over time. ${R_t^I}^*$ and ${R_t^H}^*$ denote the number of new recoveries from the I and H compartments, respectively, which are unobserved.  As written, these equations imply that infectious individuals may become hospitalized or may recover without hospitalization, and that any deaths due to the disease must first be hospitalized. While it would be possible to rewrite the equations such that infectious individuals are allowed to die without hospitalization, this introduces additional complexity to the model which is not likely to be supported by publicly available data (i.e., $D_t^*$ is observed, but ${D_t^I}^*$ and ${D_t^H}^*$ are not). Transitions between compartments are probabilistically described as
\begin{align*}
I_t^* &\sim Binom\left(S_t, \pi_t^{(SI)}\right) \\
\{H_t^*, {R_t^I}^*, I_{t + 1}\} &\sim Multinom \left(I_t, \left\{\pi^{(IH)}, \pi^{(IR)}, 1 - \pi^{(IH)} - \pi^{(IR)}\right\}\right) \\
\{D_t^*, {R_t^H}^*, H_{t + 1}\} &\sim Multinom \left(H_t, \left\{\pi^{(HD)}, \pi^{(HR)}, 1 - \pi^{(HD)} - \pi^{(HR)}\right\}\right). 
\end{align*}
The multinomial distribution is used to describe transitions from I and H as a straightforward extension of the original chain binomial SIR model. Infectious individuals may become hospitalized, recover, or stay infectious according to the probabilities $\pi^{(IH)}$, $\pi^{(IR)}$, or $1 - \pi^{(IH)} - \pi^{(IR)}$, respectively. Similarly, hospitalized individuals may die, recover, or remain hospitalized. We assume exponentially distributed lengths of time in non-susceptible compartments, leading to the following transition probabilities: $\pi^{(IH)} = 1 - \exp\left(-\lambda\right)$, $\pi^{(IR)} = 1 - \exp\left(-\gamma_1\right)$, $\pi^{(HR)} = 1 - \exp\left(-\gamma_2\right)$, $\pi^{(HD)} = 1 - \exp\left(-\phi\right)$.

The second extension to the model is to allow for the presence of undetected infections. This captures any individual who did not have a confirmed positive test reported to health officials. Undetected infections are incorporated by splitting the infectious compartment into two parts:  observed (detected) cases, $C_t$, and undetected infectious individuals, $U_t$, such that $I_t = C_t + U_t$. Correspondingly, $I_t^* = C_t^* + U_t^*$, where $C_t^*$ are the observed cases counts and $U_t^*$ is unobserved. We assume $C_t^* \sim Binom\left(I_t^*, \pi^{detect}\right)$, where $\pi^{detect}$ is the probability of an infection being detected, i.e., the case ascertainment rate. 
In many cases, there may be data from seroprevalence studies available to inform a strong prior on this detection probability. The proposed SIHRD model with undetected infections is visualized in Supplemental Figure 1. It's worth noting that we assume a constant probability of detection over time, however, in reality this may be time-varying due to changes in testing capacity  \citep{omori2020changes} or test-seeking behavior  \citep{colman2023ascertainment}. The assumption of time-invariant case ascertainment was used to better facilitate model identifiability and may be reasonable for shorter time scales such as the six to ten weeks used in our data analyses. We provide further consideration of this assumption and alternative approaches in Section \ref{disc}.

Lastly, we consider extending the alarm function of Ward et al\cite{ward2023bayesian} to allow multiple data sources to inform population behavior. The transmission probability $\pi_t^{(SI)}$ remains unchanged from Equation \ref{eq_pSI}, but we now consider the alarm to be  a function of the average observed cases and deaths over the past $m$ days, i.e., $a_t = f(C_{t-1, m}^*, D^*_{t-1, m})$, where $C^*_{t-1, m} = \frac{1}{m} \sum_{i = t-m-1}^{t-1}C^*_i$ and $D^*_{t-1, m} = \frac{1}{m} \sum_{i = t-m-1}^{t-1}D^*_i$ with smoothing parameter $m \in \{1, 2, ..., t-1\}$. While theoretically the alarm could also depend on other factors, we found that when several highly correlated data sources were used (e.g., cases, hospitalizations, \textit{and} deaths), a lack of identifiability inhibited estimation. Cases and deaths were chosen as those are most commonly available and because they are furthest apart in the disease process which reduces correlation. In a model with undetected infections the alarm could be specified as a function of the smoothed true incidence, $I_{t-1, m}^*$, however, as this is generally unknown to the population during an outbreak we find it more realistic to assume that behavior is only informed by observed data. We define the multivariable alarm as
\begin{equation} \label{eq_multialarm}
    a_t =  1 - \left[1 - \left\{\alpha C_{t-1, m}^* + (1 - \alpha)D^*_{t-1, m}\right\}/N \right]^{1/k}.
\end{equation}
This parameterization was chosen for interpretability of the $\alpha$ parameter as the relative contribution of observed cases in informing the population alarm compared to deaths. Thus, the multivariable alarm has advantages of being more flexible than a single metric alarm while also providing inference to help understand what influences population behavior. Note that the use of the multivariable alarm is not limited to the SIHRD compartmental framework, as long as data on deaths are available. However, if implemented within an SIR framework, deaths are an exogenous input as they are not generated within the model structure. In contrast, the SIHRD formulation treats both cases and deaths as endogenous inputs.

Often, the primary inferential interest in compartmental epidemic models is the reproductive number as it quantifies the magnitude of disease spread in a population. We use the approach outlined in \cite{ward2022idd} to calculate the time-varying ``effective" reproductive number, $\mathcal{R}_0(t)$ and extend this for the SIHRD model. Full details are provided in Section 2 of the Supplementary Material.

\subsection{Estimation} \label{estimation}

Inference is performed in the Bayesian framework. The log-likelihood for the SIR model with IDD transmission is $\ell(\mathbf{I}^*, \mathbf{R}^* | \boldsymbol{\Theta}) = \sum_{t = 0}^{\tau}  \left\{\log \binom{S_t}{I^*_t} + I^*_t \log \pi_t^{(SI)} + \left(S_t-I^*_t \right) \log \left(1-\pi_t^{(SI)} \right) \right\}$, with parameter vector $\Theta$ containing $\beta$, $w_0$, $\nu$ and any parameters associated with an alarm function (e.g., $k$ and $\alpha$ for the multivariable alarm). Commonly, $\beta$ and $k$ are given vague gamma priors (e.g., Gamma(0.1, 0.1)), and informative priors are used for the parameters $w_0$ and $\nu$ which describe the logistic decay curve of transmissibility over an individual's infection. As $w_0$ is the inflection point it can be thought of as the average length of time individuals are infectious using a normal prior with small variance. For the decay rate, $\nu$, we typically use Gamma priors centered on a value of 1 (e.g., Gamma(100, 100)) to enforce a sharp decrease in transmission around the inflection point for interpretability of $w_0$. The implication of this prior specification is illustrated in Supplementary Figure S2

For the proposed SIHRD model with undetected infections the complete log-likelihood is 
\begin{align*} 
&\ell(\mathbf{I}^*, \mathbf{C}^*, \mathbf{H}^*, \mathbf{R^I}^*, \mathbf{R^H}^*, \mathbf{D}^* | \boldsymbol{\Theta}) = \\
&\hspace{9mm} \sum_{t = 0}^{\tau}  
    \Big\{\log \binom{S_t}{I^*_t} + I^*_t \log \pi_t^{(SI)} + \left(S_t-I^*_t \right) \log \left(1-\pi_t^{(SI)} \right)\\
&\hspace{9mm} + \log \frac{I_t!}{H^*_t!{R^I_t}^*!I_{t+1}!} + H^*_t \log \pi^{(IH)} 
 + {R^I_t}^* \log \pi^{(IR)} + I_{t + 1} \log \left(1- \pi^{(IH)} - \pi^{(IR)} \right) \\
&\hspace{9mm} + \log \frac{H_t!}{D^*_t!{R^H_t}^*!H_{t+1}!} + D^*_t \log \pi^{(HD)} 
 + {R^H_t}^* \log \pi^{(HR)} + H_{t + 1} \log \left(1- \pi^{(HD)} - \pi^{(HR)} \right) \\
 &\hspace{9mm} + \log \binom{I^*_t}{C^*_t} + C^*_t \log \pi^{detect} + \left(I^*_t-C^*_t \right) \log \left(1-\pi^{detect} \right) \Big\},
\end{align*}
with parameter vector $\Theta$ containing $\beta$, $\gamma_1$, $\gamma_2$, $\lambda$, $\phi$, $\pi^{detect}$, $k$ and $\alpha$ for the model with the multivariable alarm function. Complete data would provide $\mathbf{I}^*$, $\mathbf{C}^*$, $\mathbf{H}^*$, $\mathbf{R^I}^*$, $\mathbf{R^H}^*$, and $\mathbf{D}^*$ and initial conditions ($S_0$, $I_0$, $H_0$, $R_0$, $D_0$), however, from publicly available data we typically only observe $\mathbf{C}^*$, $\mathbf{D}^*$, and sometimes $\mathbf{H}^*$. We assume these three data sources are available and use the data-augmented MCMC approach of Lekone and Finkenst\"adt \citep{lekone} to impute $\mathbf{R^I}^*$ and $\mathbf{R^H}^*$. We are not able to avoid their imputation through the IDD approach taken in the SIR model due to the multinomial transitions from the I and H compartments as the IDD specification has not yet been extended for this purpose. Data augmentation is also used to estimate the unobserved true incidence $\mathbf{I}^*$ in models that incorporate undetected infections. When undetected infections are ignored it is assumed that $\mathbf{I}^* = \mathbf{C}^*$ and the last component of the likelihood above is not included.

As in the SIR model, vague Gamma priors are used for $\beta$ and $k$ in the SIHRD model as is common in practice \citep{ward2023bayesian, ward2025framework}. Informative Gamma priors may be used for the parameters $\gamma_1$ (I to R), $\gamma_2$ (H to R), $\lambda$ (I to H), and $\phi$ (H to D) in the presence of prior knowledge about the disease state. For imputing $\mathbf{I}^*$, strong priors are required for $\gamma_1$ and $\lambda$ to induce identifiability. Otherwise, the same overall reproductive number could be produced from a high transmission rate and short infectious periods or a relatively lower transmission rate with long infectious periods. When hospitalizations and deaths are observed, less informative priors can be used for $\lambda$ and $\phi$ as data is available to estimate these parameters. However, more informative priors on $\gamma_1$ and $\gamma_2$ can greatly aid in estimation and are often available based on the interpretation of $1/\gamma_1$ ($1/\gamma_2$) as the average length of time spent infectious (hospitalized) before recovery. We assign $\pi^{detect}$ a strong Beta prior centered on the mean used to define $\mathbf{I}^*$. We consider the specification, Beta($a\times \rho$, $(1-a)\times \rho$), which has mean $a$ and variance $\frac{a(1-a)}{\rho+1}$ so $\rho$ can be treated as a scaling parameter where larger values lead to reduced variance. We consider $\rho = 10000$ in our simulation study and data applications as a strong prior is required for identifiability. The relative importance parameter $\alpha$ is given a Uniform(0, 1) prior as our interest lies in estimating this from observed data without any strong prior assumptions.

All models were implemented using the R package \texttt{nimble}  \citep{nimble-article:2017, nimble-software:2021} with user-defined samplers for the data-augmented MCMC algorithms. For all analyses, three MCMC chains were run using various starting values of the parameters and convergence was assessed by ensuring the Gelman and Rubin diagnostic value for all parameters fell below 1.1  \citep{gelmanrubin1992}. All code to reproduce the simulations and data analyses presented is available at \url{https://github.com/ceward18/multipleDataBCM}.

\subsection{Assessing Model Fit} \label{modelfit}

We evaluate model fit using the posterior predictive distribution over the time period used in model fitting. This is computed by taking a large number of draws (e.g., 10,000) from the posterior distribution of all model parameters and using each draw to simulate epidemics through forward simulation according to the specified model starting from the initial conditions. We summarize the posterior predictive distribution of the epidemic over time using the mean and 95\% credible intervals of the simulated epidemic trajectories. We then compare the posterior predictive distribution to the observed data visually, with more overlap between the posterior predictive distribution and the observed data indicating better model fit. For SIR models, the observed data is the single time series of case counts, whereas for the SIHRD models, the observed data also includes time series of hospitalizations and deaths. For models with behavioral change captured in an alarm function, the alarm is dynamically updated as case and/or death counts are simulated for each posterior predicted epidemic. However, this becomes problematic for an SIR model with cases and deaths informing the alarm function as deaths are not part of the model and are subsequently not simulated as part of the posterior predictive process. Posterior prediction is only possible when the inputs to the alarm function are included in the compartmental structure, and is thus not possible for an SIR model with a multivariable alarm depending on both cases and deaths.

In addition to the visual comparisons using the posterior predictive distributions, we also use the Widely Applicable Information Criteria (WAIC) \citep{watanabe2010asymptotic}. WAIC is a likelihood-based estimate of prediction error which includes a penalty for the effective number of model parameters \citep{gelman2013bayesian}. As presented in Section \ref{estimation}, the SIR and SIHRD models have different likelihoods which means the WAIC values are not comparable across the two compartmental structures. However, within each compartmental structure, WAIC can still be used to determine the best specification of the alarm function \citep{ward2023bayesian}.

\section{Simulation Study} \label{simStudy}

\subsection{Simulation Set-up}

\begin{figure}[ht!]
    \centering
    \includegraphics[width = 0.9\textwidth, trim = {0 1cm 0 0}, clip]{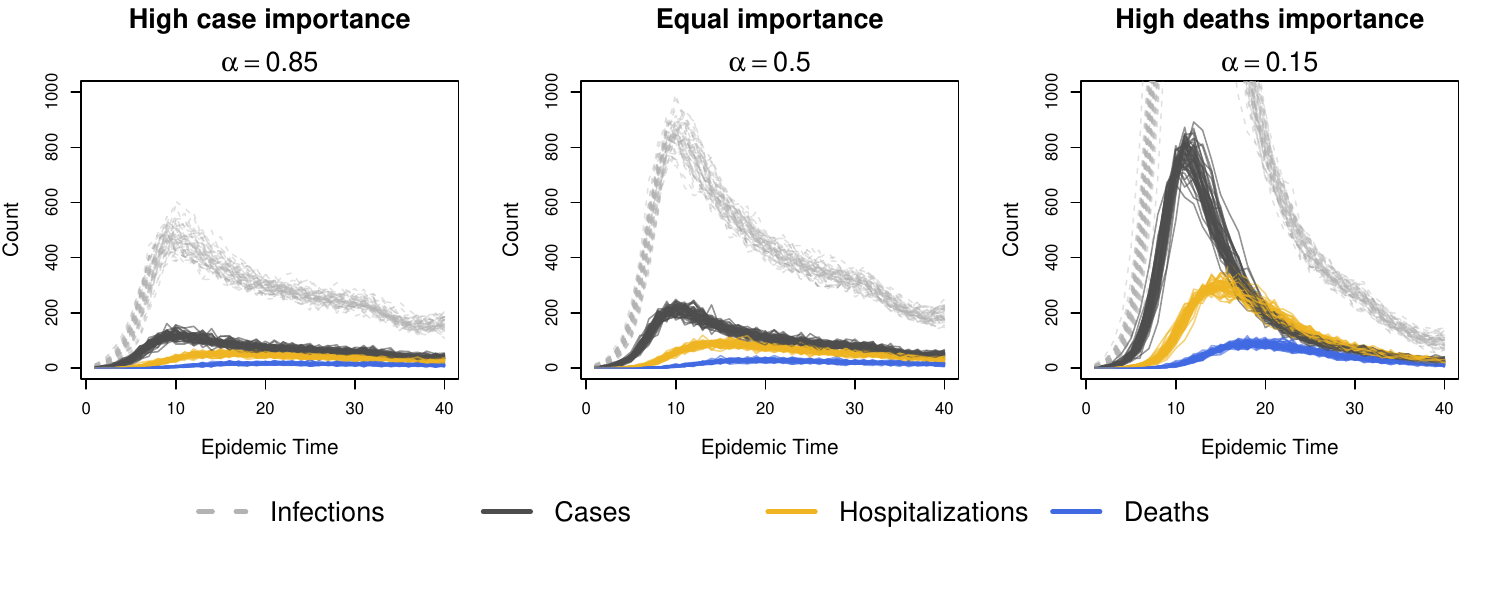}
    \caption{Simulated epidemics from three scenarios capturing various importance of cases and deaths informing  behavioral change. Plotted over time are the total (unobserved) infections (dashed gray), observed cases (solid black), hospitalizations (solid gold), and deaths (solid blue). In the high deaths importance scenario, infections peak at counts around 3,000 but this was excluded from the figure to highlight more subtle differences between high case importance and equal importance generating scenarios.}
    \label{fig_simcurves}
\end{figure}

A simulation study was designed to achieve the following objectives: 1) understand how estimation is impacted if the presence of undetected cases is ignored, 2) determine whether the full SIHRD model is necessary or if a simpler SIR model can still provide the same inferential results, and 3) ascertain if the multivariable alarm is estimable and how it affects model fit compared to a univariate alarm or ignoring behavioral change completely. To achieve these goals, epidemics were simulated under the full SIHRD model with undetected infections and alarm based on smoothed cases and deaths as specified in Equation \ref{eq_multialarm} with smoothing over the past $m = 30$ days. All epidemics were simulated in a population of one million people with five initially infected individuals and were recorded for 40 days. The detection probability was set to $\pi^{detect} = $ 25\% for all simulations. Three values of the weight parameter $\alpha$ in the alarm function were used for data generation: $\alpha = 0.85$ for high case importance,  $\alpha = 0.5$ for equal importance of cases and deaths, and $\alpha = 0.15$ for high death importance. 50 epidemics were simulated in each of the three data generating scenarios. All other model parameters were given the same value across the three  scenarios (full specifications are provided in Supplementary Table 1). This led to epidemics as illustrated in Figure \ref{fig_simcurves}.

Six models were fitted to each simulated epidemic to enumerate all possible combinations of the two compartmental structures (SIHRD and SIR) and three alarm specifications: multivariable alarm based on cases and deaths, alarm based on cases only as specified in Equation \ref{eq_unialarm}, and no alarm ($a_t = 0$ for all $t$). Each of these models was fit under two assumptions: observed cases are the true incidence (ignoring undetected infections) and allowing for undetected infections to be incorporated in the modeling process as specified in Section \ref{estimation}. When undetected infections were modeled, we assume a correctly specified prior, such that $E(\pi^{detect}) = 0.25$ with small variance. Other priors were specified as in Section \ref{estimation}, and a full description is provided in Supplementary Table 2. Three MCMC chains with different initial values were run for each model with a burn-in of 100,000 iterations and 300,000 post burn-in samples drawn every 10th iteration. Convergence was established by a Gelman and Rubin diagnostic value below 1.1  \citep{gelmanrubin1992}. A small number of the SIHRD models ($<$ 1\%) did not converge and have been excluded from the results.

\subsection{Simulation Results}

\begin{figure}[ht!]
    \centering
    \includegraphics[width = 0.95\textwidth]{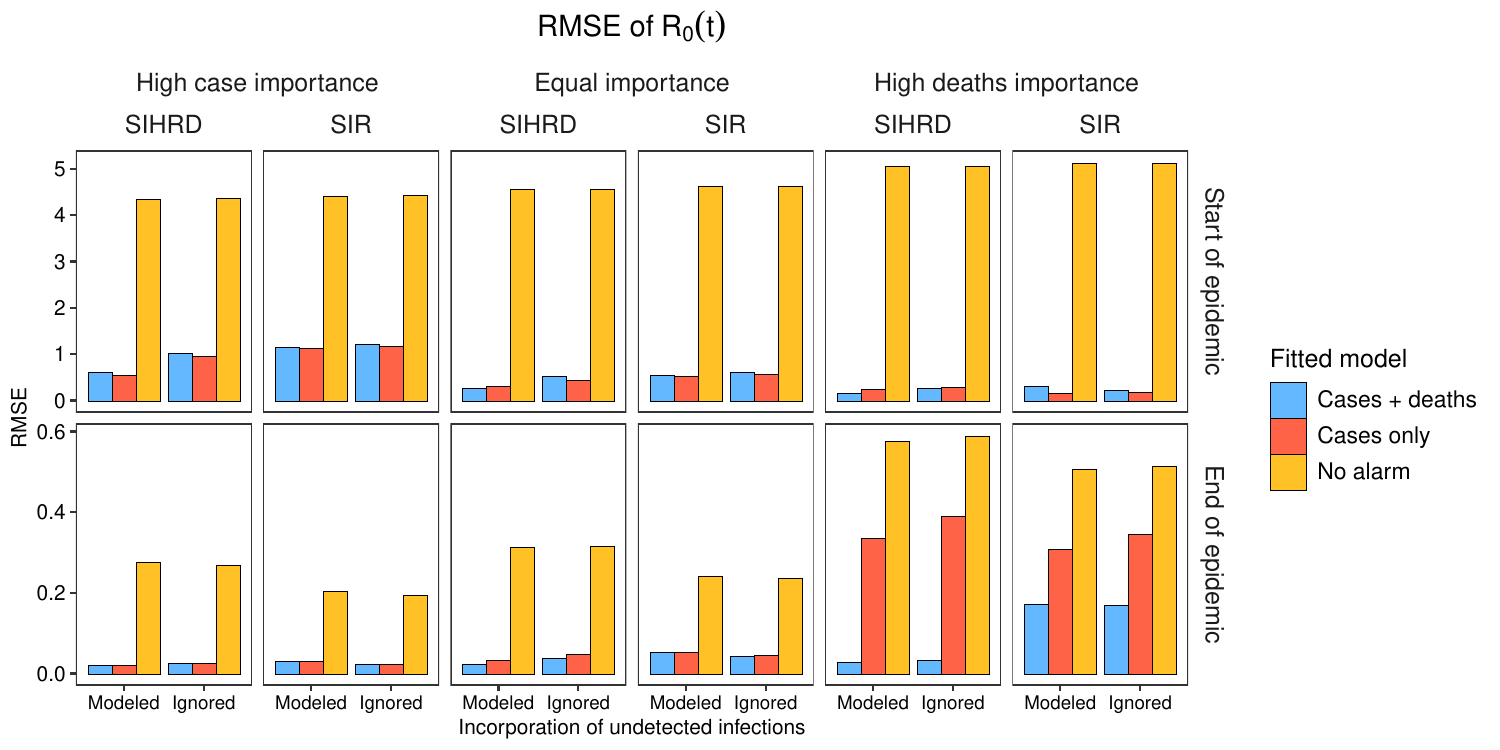}
    \caption{Root mean squared error (RMSE) of the effective reproductive number, $\mathcal{R}_0(t)$, at the start and end of the simulated epidemics for the three data generating scenarios, the SIHRD and SIR models, and modeling or ignoring undetected infections. }
    \label{fig:sim_r0RMSE}
\end{figure}

We use estimation of the reproductive number to assess the impact of ignoring undetected cases and to compare the SIHRD and SIR compartmental structures due to its epidemiological importance. Across simulations $k = 1, ..., K$, each model's estimated $\hat{\mathcal{R}}_0(t)$ was computed and compared to the true $\mathcal{R}_0(t)$ using the root mean squared error (RMSE) computed as $\sqrt{\frac{1}{K}\sum_{k = 1}^K \left\{ \hat{\mathcal{R}}_0(t) - \mathcal{R}_0(t)\right\}^2}$ at each time point $t$. In Figure \ref{fig:sim_r0RMSE}, we present RMSE at the first and last time point to illustrate estimation prior to behavioral change occurring (start) and after behavioral change has reduced transmission (end). RMSE and the posterior distribution of $\mathcal{R}_0(t)$ from one simulation across all time points are provided in Supplementary Figures 1 and 2.

First, we find that models that do not incorporate any behavioral change have very poor estimation of transmission at both the start and end of the epidemics, as indicated by the large RMSE values for the models with no alarm. In assessing estimation of the reproductive number when the presence of undetected cases is ignored, we observe only minimal differences in RMSE of $\mathcal{R}_0(t)$. This is consistent with previous literature indicating that the estimation of the reproductive number is unaffected by under-detection when there is a constant detection probability  \citep{cori2013new, gostic2020practical}. Inspecting the posterior distributions of all parameters in the SIHRD models (Supplementary Figures 4-6), we find that while the transmission parameters ($\beta$, $k$, $\alpha$) were generally similar in distribution whether or not undetected infections were incorporated into the modeling, the parameters associated with the length of the infectious period were biased such that $\pi^{(IH)}$ was larger and $\pi^{(IR)}$ was smaller than their true values. Since hospitalizations over time are observed, ignoring undetected infections increases $\pi^{(IH)}$ as there are less infected people available to be hospitalized. In combination, the tradeoff between these two probabilities ends up having minimal impact on the estimation of $\mathcal{R}_0(t)$. Evaluating the posterior distributions of SIR model parameters (Supplementary Figures 7-9), we see minimal changes in the posterior distributions when undetected infections were incorporated or ignored. Thus any differences in the reproductive number would be a result of differences in the number of susceptibles over time. However, as the simulated epidemics are infecting a relatively low proportion of individuals in a large population, incorporating undetected infections has only a minor impact on the total number of susceptible individuals and therefore on the estimated $\hat{\mathcal{R}}_0(t)$. Since there are minimal differences in estimation between models that do and do not incorporate undetected infections, we focus the remaining results on the more flexible model where undetected infectious are incorporated into the model structure.

We also evaluate Figure \ref{fig:sim_r0RMSE} for differences in the RMSE of $\mathcal{R}_0(t)$ between the two compartmental structures and the full multivariable alarm compared to the reduced alarm based on cases only. For both modeling choices, we find that there is little penalty in using the simpler model or alarm structure in terms of estimation of $\mathcal{R}_0(t)$ in the high case importance and equal importance data generating scenarios. Where it becomes most important to have the correct model specification is in the scenario where deaths are highly important in driving behavior and one is interested in estimating $\mathcal{R}_0(t)$ towards the end of the epidemic after behavioral changes have occurred. It's intuitive that a model ignoring deaths (i.e., the cases only model) would do poorly in this scenario where deaths are highly important. It is perhaps more interesting that even when the alarm function was correctly specified to include both cases and deaths, if the SIR compartmental structure was used  $\mathcal{R}_0$ estimation suffers. Since we observe substantial differences in estimation of $\mathcal{R}_0(t)$ between the SIR and SIHRD models only in settings where deaths have a strong influence on behavior, this suggests that the improved performance of the SIHRD model is not simply due to the more detailed compartmental structure. Rather, it arises from the treatment of deaths as an endogenous input to the alarm function. These results indicate that when an outcome strongly influences behavior, representing it endogenously within the model may be important for obtaining accurate estimation of epidemiological quantities such as $\mathcal{R}_0(t)$.

\begin{figure}[ht!]
    \centering
    \includegraphics[width = 0.9\textwidth]{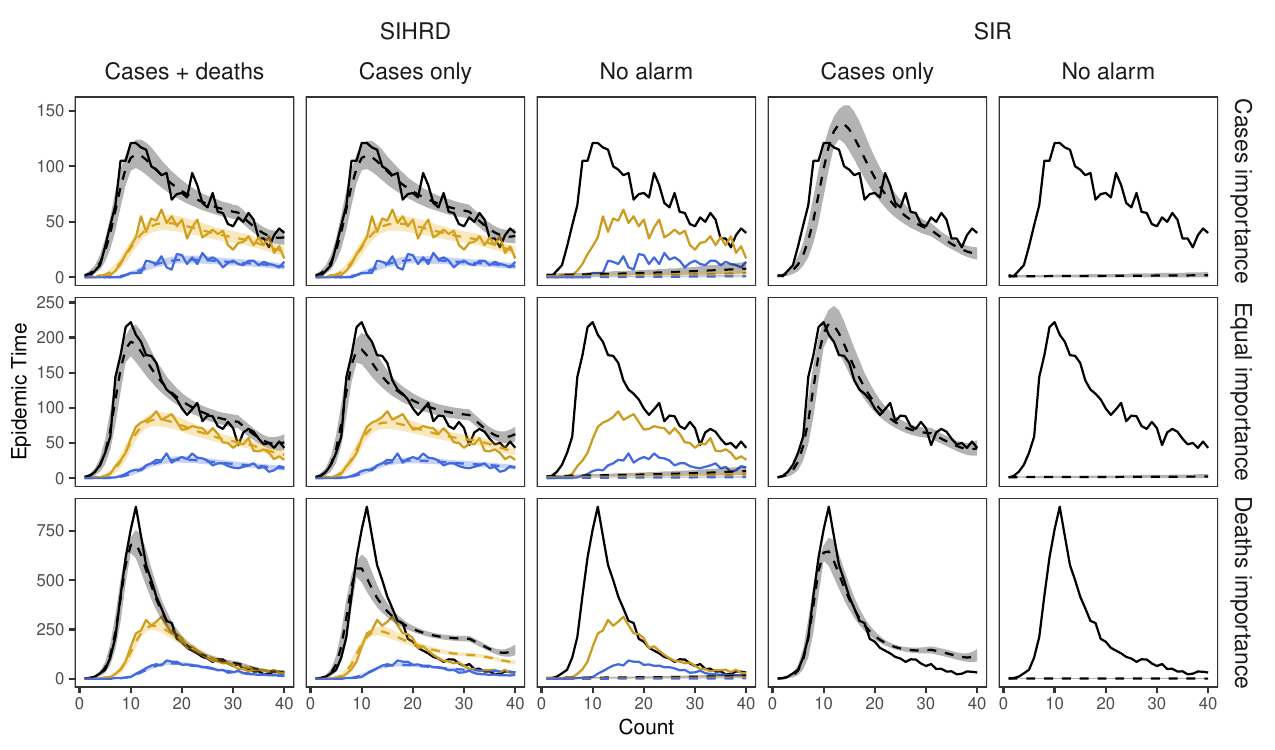}
    \caption{Posterior predictive means (dashed lines) and 95\% credible intervals (shaded regions) for one randomly selected simulation compared to true cases (solid black), hospitalizations (solid gold), and deaths (solid blue). These results are from models that incorporate undetected infections.}
    \label{fig:sim_postPred}
\end{figure}

Next, we evaluate model fit using the posterior predictive distributions as described in Section \ref{modelfit}. The posterior predictive distributions for one randomly selected simulated epidemic from each data generating scenarios are presented in Figure \ref{fig:sim_postPred}. 
As we would expect given the poor estimation of $\mathcal{R}_0(t)$ for models that do not incorporate behavioral change, we find that these models also fit very poorly. Transmission is severely underestimated leading to posterior predictive distributions where on average less than 10 individuals are infected across both the SIHRD and SIR compartmental structures and all three data generating scenarios. Models ignoring behavioral change must estimate an average transmission rate across epidemics with a brief increase in cases followed by a longer period of declining cases, and therefore estimate $\mathcal{R}_0$ right around 1 across time (Supplementary Figure 4). Also as expected, the posterior predictive distributions for the SIHRD model with the full multivariable alarm are fairly close to the truth across all three data generating scenarios. When the alarm is specified based on cases only, the SIHRD model fits well in scenarios of high case importance or equal importance but significantly underestimates the peak and overestimates the tail when deaths are highly important. Similar trends are seen for the SIR model with cases only in the alarm function, and this model also seems to overestimate the peak in the high case importance scenario.

\begin{figure}[ht!]
    \centering
    \includegraphics[width = 0.9\textwidth]{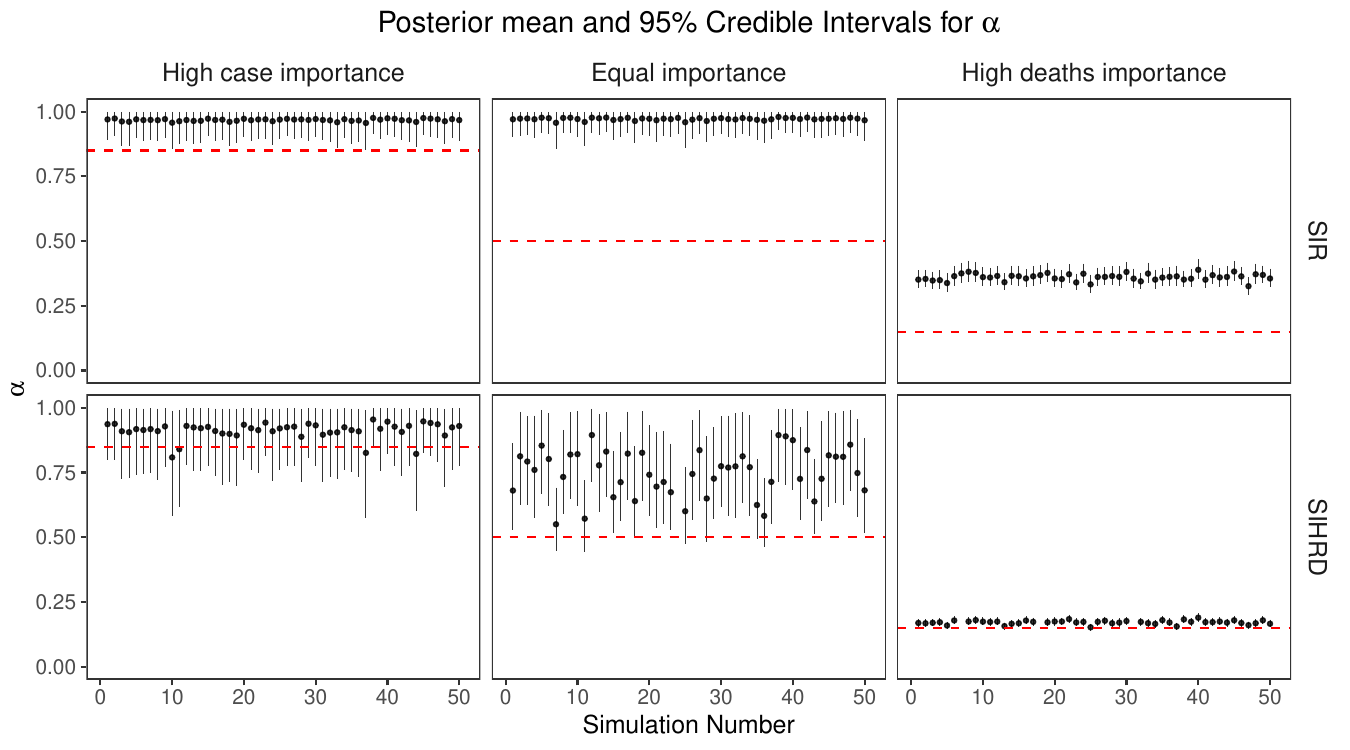}
    \caption{Posterior means and 95\% credible intervals for $\alpha$ across all simulations for the three data generating scenarios (columns) and the SIHRD and SIR models with multivariable behavioral change as specified in Equation \ref{eq_multialarm} (rows). The true parameters used in simulation are shown by the dashed red lines. These results are from models that incorporate undetected infections.}
    \label{fig:sim_alphas}
\end{figure}

Finally, to evaluate the estimation of the multivariable alarm we focus on estimation of $\alpha$, the relative importance of cases in the alarm function across both the SIHRD and SIR models for all simulations (Figure \ref{fig:sim_alphas}). These results illustrate that the SIR model is biased towards cases being more important, whereas the SIHRD model does a much better job of capturing this parameter, particularly in the high deaths importance and equal importance scenarios. This is also likely the source of the increase in RMSE of $\mathcal{R}_0$ for the SIR model in the high deaths importance scenario observed in Figure \ref{fig:sim_r0RMSE}. We hypothesize that this may be due to the fact that deaths are directly included into the compartmental structure in the SIHRD model. The SIHRD model does still show a small amount of bias towards higher case importance as reflected by point estimates being consistently above the true value. The estimation of $\alpha$ is worst in the equal importance scenario, but this is not necessarily surprising as we see that epidemics generated under high case importance and equal importance scenarios are more similar than epidemics generated under the high deaths importance scenario (Figure \ref{fig_simcurves}) which may indicate that it is harder to differentiate between moderate and high values of $\alpha$ than between moderate and low values of $\alpha$. 

\section{Data Analysis} \label{analysis}

\subsection{COVID-19 in Miami and Montr\'{e}al}

\begin{figure}[ht!]
    \centering
    \includegraphics[width = 0.9\textwidth]{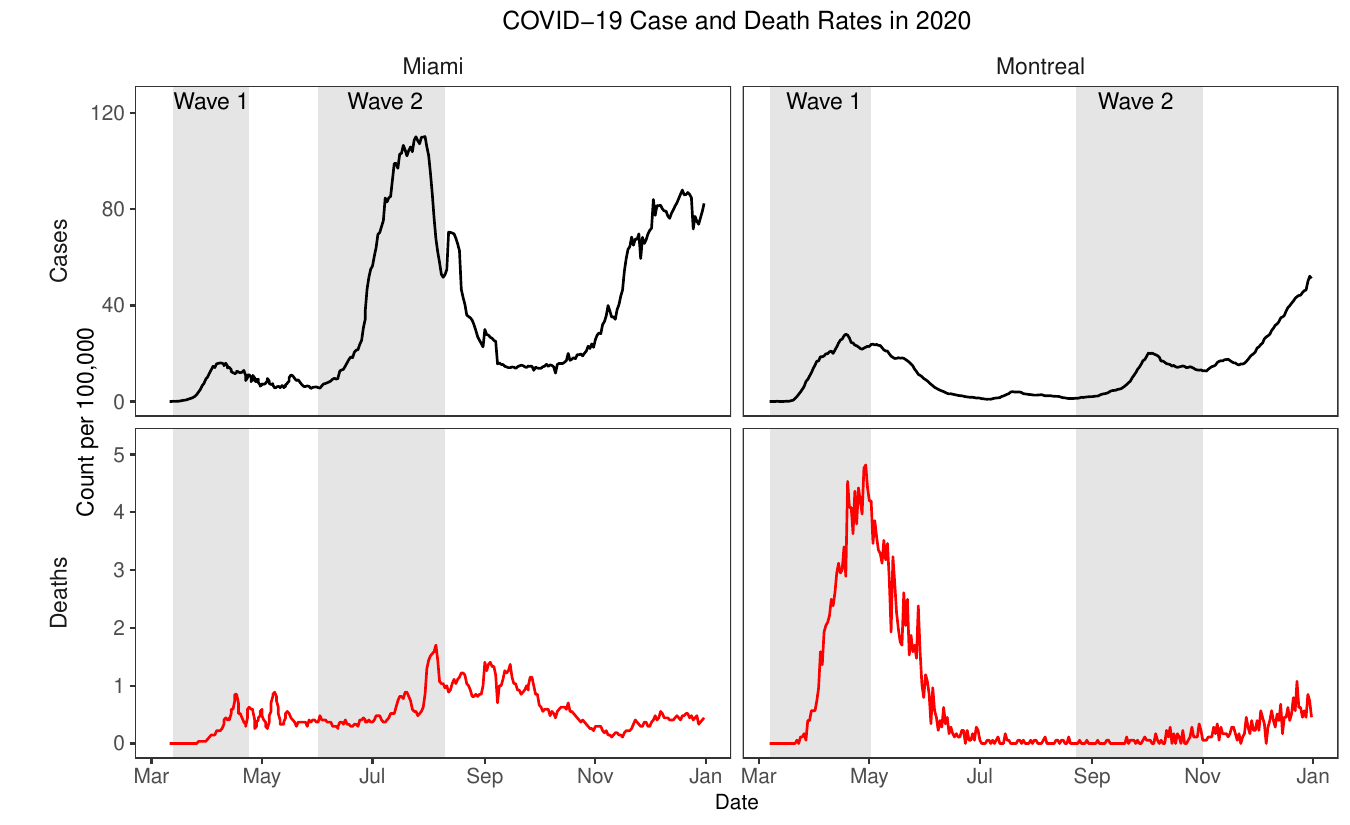}
    \caption{COVID-19 case and death rates per 100,000 during 2020 in Miami-Dade County, Florida and Montr\'{e}al, Qu\'{e}bec. Gray shaded regions are those used in model fitting. In Miami, Wave 1 started on March 11, 2020 and Wave 2 started on June 1. In Montr\'{e}al, Wave 1 started on March 5, 2020 and Wave 2 started on August 23, 2020.}
    \label{fig:observed_data}
\end{figure}

To illustrate the model utility in practice, we analyzed data from the first two waves of COVID-19 in Miami-Dade County (Miami), Florida in the United States and Montr\'{e}al, Qu\'{e}bec in Canada. These locations were chosen as both were severely affected by COVID-19, with Montr\'{e}al having the highest death rate in Canada and Miami-Dade County having the third highest case rate in the United States (NYT COVID tracker). These two locations also provide an interesting comparative case study as they have very different trends in case and death rates over time. As illustrated by Figure \ref{fig:observed_data}, while Miami saw a significant increase in cases and a small increase in deaths in the second wave, Montr\'{e}al had similar case rates during both waves, but a large decrease in the death rate during the second wave. Based on these trends, we hypothesize that the relative importance of cases and deaths will be different between the two cities and over time within each city.

Case and death data for Miami-Dade county was obtained from the New York Times, based on reports from state and local health agencies  \citep{NYtimesdata}. Miami hospitalization data was obtained from the Miami-Dade COVID Project  \citep{williams2021lessons, miamicovidproject} and provides the number of COVID-19 positive patients admitted on each day. The counts of newcase and death counts start on March 11, 2020 when the first case occurred in the county, however hospitalization data is not available until April 2, 2020. The missing hospitalization data was imputed using data augmentation. The population size was set at $N =$ 2,701,767, the value reported by the 2020 U.S. census. Daily counts of new cases, hospital admissions, and deaths for Montr\'{e}al were obtained from the National Institute of Public Health of Qu\'{e}bec  \citep{montrealData}, with the first positive case reported on March 5, 2020. The population size for Montr\'{e}al was set to $N =$ 1,762,949 as this was the value reported in the 2021 Canadian Census of Population.

To illustrate the use of these models during an active outbreak, models were fit to the first six to ten weeks of each wave. The amount of data included varied by city and wave, but was chosen to include only a few weeks of data after the peak. Sensitivity analyses using different numbers of weeks of data in model fitting are presented in Supplementary Table 5. As in the simulation study, six models were fitted to encompass both compartmental structures and three alarm specifications. Alarms were based on smoothed cases and deaths over the past $m = 30$ days. Despite the simulation results indicating minimal benefit of incorporating undetected infections, all models incorporated this uncertainty. Prior distributions were specified as in Section \ref{estimation} and are detailed in Supplementary Tables 3 and 4. Priors for the probability of detection in Miami were specified based on the Institute for Health Metrics and Evaluation (IHME) estimates of total infections and reported cases in Florida during the corresponding waves, leading to prior means of 12\% in wave 1 and 34\% in wave 2  \citep{ihme_miami}. For Montr\'{e}al, these priors were based on IHME COVID-19 policy briefing on the percent of COVID-19 infections detected across Canada, with prior means of 25\% detection in wave 1 and 40\% detection in wave 2  \citep{ihme_canada}. Priors for the removal rate for hospitalized patients were chosen to reflect an average length of stay of 15 days  \citep{alimohamadi2022hospital} and priors for the removal rate and IDD curve for infected but not hospitalized individuals were chosen to reflect an average seven day infectious period before recovery. The initial values in each compartment, $\{S_0, I_0, H_0, R_0, D_0\}$ or $\{S_0, I_0, R_0\}$, were estimated using strong priors based on the previous epidemic trajectory. Specific details on the specification are provided in Section 5.1 of the Supplementary Material. As in the simulation study, three chains were run and convergence established by a Gelman Rubin diagnostic value below 1.1.

\subsection{Results}

First, we evaluate model fit using WAIC and compare the estimates of the relative importance of cases ($\alpha$) in informing the alarm for both cities using the SIHRD and SIR models (Table \ref{tab:data_post_alpha_waic}). We find that, in general, estimates of $\alpha$ are similar between the two approaches, but the SIHRD model tends to attribute more behavioral change to deaths than the SIR model, as we would expect based on the results of our simulation study. During the first wave of COVID-19 cases in both cities, the estimated relative importance parameter $\alpha$ in the multivariable alarm model indicates that the majority  ($\sim$88-94\%) of the population alarm is due to cases during the first wave. This concurs with the WAIC values, which are essentially equivalent between the multivariable alarm and the cases only alarm in both cities. However, in Wave 2 in Miami we found that deaths were highly influential as the estimated $\hat{\alpha}$ values are close to zero and WAIC shows that model fit is significantly improved through the addition of deaths to the alarm function for both the SIHRD and SIR models. This finding concurs with the observed data, where the case counts rapidly rise while the death counts remained relatively stable until mid-July, and only after that do the case counts start to come down. This indicates that the timing of the behavioral change better coincides with increasing deaths counts rather than increasing case counts. In Wave 2 in Montr\'{e}al, the models estimate relatively equal contribution of case and deaths inform alarm, however the credible intervals are extremely wide. These wide intervals are likely due to the low number of deaths during the time period of model fitting (Figure \ref{fig:observed_data}). Again the WAIC values differ by $<1$ point between the cases and deaths alarm and the cases only alarm model. Overall the WAIC values indicate that there is little downside in terms of model fit for incorporating deaths into the alarm function even when they are not very important, and that model fit is much worse when behavioral change is not incorporated at all.

\begin{table}[]
    \centering
    \small
\begin{tabular}{lllcccc}
\toprule
\multicolumn{3}{c}{\textbf{ }} & \multicolumn{2}{c}{\textbf{SIHRD}} & \multicolumn{2}{c}{\textbf{SIR}} \\
\cmidrule(l{3pt}r{3pt}){4-5} \cmidrule(l{3pt}r{3pt}){6-7}
\textbf{City} & \textbf{Wave} & \textbf{Model fitted} & \textbf{WAIC} & $\hat \alpha$ & \textbf{WAIC} & $\hat \alpha$\\
\midrule
 &  & Cases + deaths & 627.0 & 0.875 (0.619, 0.996) & 516.2 & 0.881 (0.636, 0.996)\\

 &  & Cases only & \cellcolor{gray!25}625.2 & - & \cellcolor{gray!25}515.1 & -\\

 & \multirow{-3}{*}{\raggedright\arraybackslash Wave 1} & No alarm & 1723.0 & - & 1141.6 & -\\
\cmidrule{2-7}
 &  & Cases + deaths & \cellcolor{gray!25}1649.1 & 0.000 (0.000, 0.000) & \cellcolor{gray!25}1132.8 & 0.002 (0.001, 0.003)\\

 &  & Cases only & 2156.0 & - & 1391.0 & -\\

\multirow{-6}{*}{\raggedright\arraybackslash Miami} & \multirow{-3}{*}{\raggedright\arraybackslash Wave 2} & No alarm & 3596.5 & - & 1420.8 & -\\
\cmidrule{1-7}
 &  & Cases + deaths & \cellcolor{gray!25}2454.3 & 0.941 (0.800, 0.998) & 739.3 & 0.942 (0.805, 0.998)\\

 &  & Cases only & 2454.5 & - & \cellcolor{gray!25}738.6 & -\\

 & \multirow{-3}{*}{\raggedright\arraybackslash Wave 1} & No alarm & 3006.7 & - & 875.2 & -\\
\cmidrule{2-7}
 &  & Cases + deaths & 1349.8 & 0.456 (0.023, 0.972) & 576.1 & 0.616 (0.149, 0.983)\\

 &  & Cases only & \cellcolor{gray!25}1349.3 & - & \cellcolor{gray!25}575.6 & -\\

\multirow{-6}{*}{\raggedright\arraybackslash  Montr\'{e}al} & \multirow{-3}{*}{\raggedright\arraybackslash Wave 2} & No alarm & 1520.8 & - & 647.8 & -\\
\bottomrule
\end{tabular}
    \caption{ WAIC values for all models and posterior means (95\% credible intervals) of $\alpha$ for multivariable alarm models. SIHRD and SIR models have different likelihoods so the WAIC values are not comparable across compartmental structures. Lowest WAIC within each wave/compartment grouping is shaded. The parameter $\alpha$  provides the relative importance of cases in informing the population alarm, with values closer to one indicating high case importance and values closer to zero indicating high death importance. }
    \label{tab:data_post_alpha_waic}
\end{table}

\begin{figure}[ht!]
    \centering
    \includegraphics[width = 0.95\textwidth]{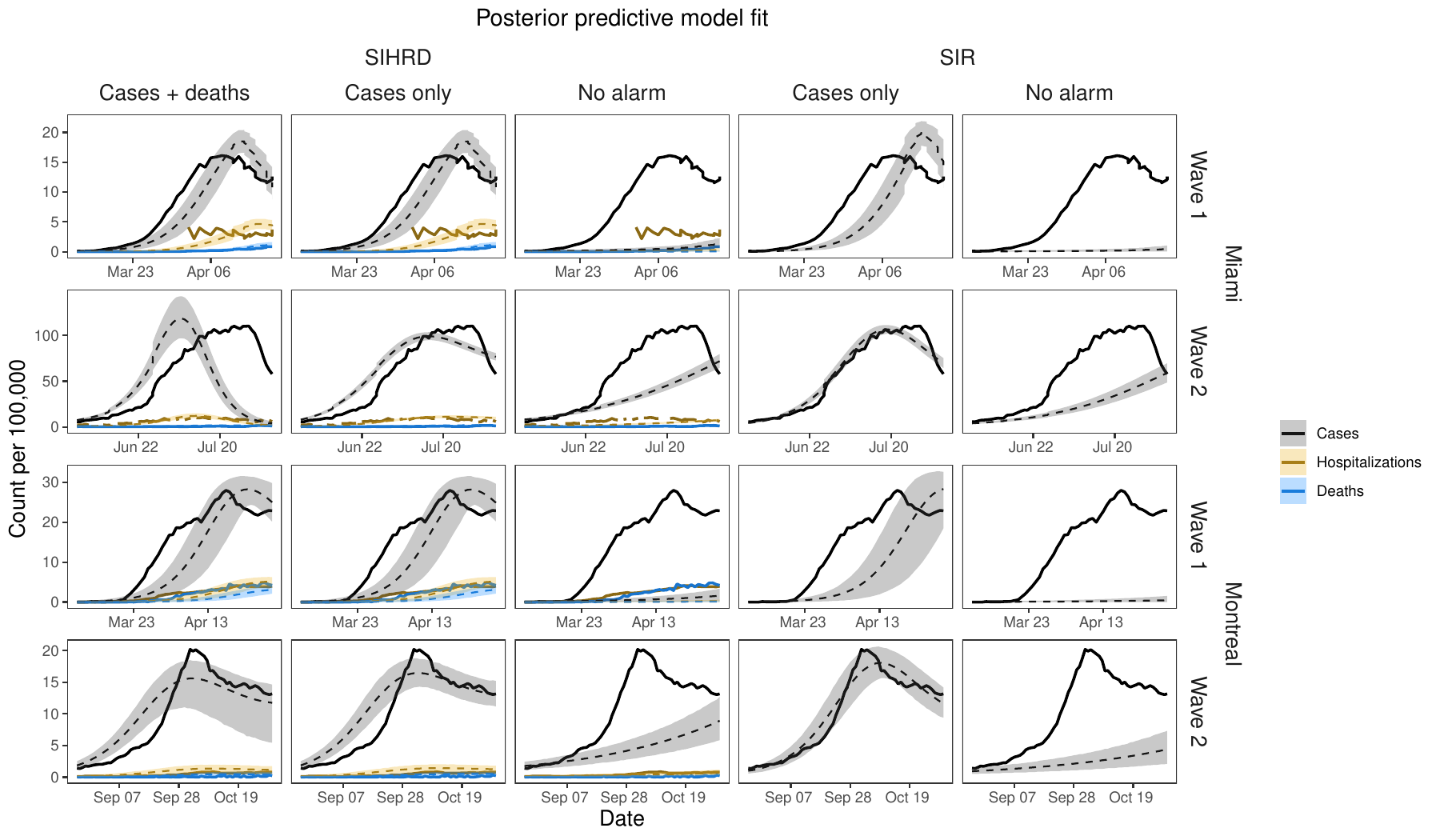}
    \caption{Posterior predictive means (dashed lines) and 95\% credible intervals (shaded regions) for cases, hospitalizations, and deaths compared to observed cases (solid black), hospitalizations (solid gold), and deaths (solid blue) for two COVID-19 waves in Miami and Montr\'{e}al. Hospitalizations and deaths are only shown for the SIHRD model as posterior prediction of deaths is not possible from the SIR model (see Section \ref{modelfit}). Hospitalization data for Wave 1 in Miami was only available from April 2, 2020 and was otherwise imputed. }
    \label{fig:data_postpred}
\end{figure}

As WAIC can only provide insight into relative specification of the alarm function within a compartmental structure, we also use the posterior predictive distribution of cases (and hospitalizations and deaths for the SIHRD models) trajectories as a global comparison of model fit across all models. (Figure \ref{fig:data_postpred}). As expected, the posterior predictive distributions of the models with no alarm do not resemble the observed data as transmission is severely underestimated from being averaged across periods of both increasing and decreasing case counts. Comparing the SIHRD models to the SIR models, we find the SIHRD models provide better fit to Wave 1 in both cities, although the posterior prediction of cases for the SIHRD models still tends to underestimate the initial increase in cases and estimate a slightly delayed peak. In Wave 2, the SIR cases only model has the most overlap between the posterior predictive distribution and the observed case counts. This is not surprising as our simulation study also found this model to have adequate posterior predictive fit until later into the wave even when it was not the true data generating model. Of note, the posterior predictive distribution of cases in Wave 2 in Miami for the SIHRD model with cases and deaths in the alarm function does not match the observed data, despite having a significantly lower WAIC value then the SIHRD model with only cases informing the alarm. Additionally for this wave, the SIR cases only model has a posterior predictive distribution that best matches the observed data. Although the posterior predictive distribution of cases was poor, both SIHRD models had reasonable posterior predictive coverage of the observed hospitalization and death curves (see Supplementary Figure 10 for the zoomed in posterior predictive distributions of hospitalizations and deaths). Combined, these results may indicate that the strong prior on the constant hospitalization rate, which is needed for incorporating the undetected infections is misspecified. Further discussion on this assumption and the feasibility of potential alternate specifications is provided in Section \ref{disc}.

\begin{figure}[ht!]
    \centering
    \includegraphics[width = 0.95\textwidth]{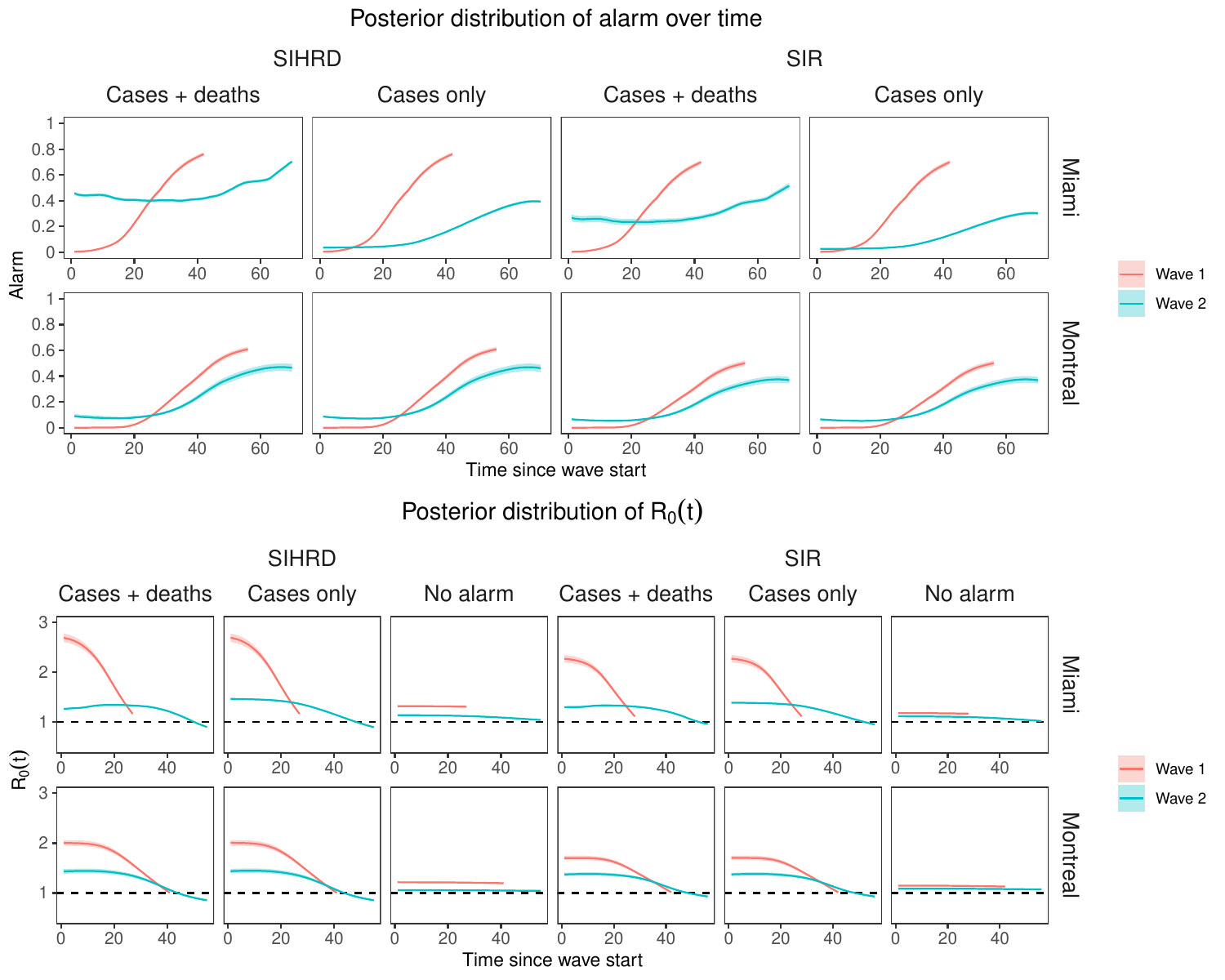}
    \caption{Estimated time-varying reproductive numbers (top) and posterior predictive epidemic trajectory (bottom) for both cities. Solid lines represent the means and the shaded regions represent the 95\% credible intervals. Estimates for Wave 1 are shown in red and estimates for Wave 2 are shown in blue.}
    \label{fig:data_r0}
\end{figure}

Lastly, we evaluate the posterior distributions of the alarm functions and reproductive numbers over time, which illustrates several other important findings (Figure \ref{fig:data_r0}). In both cities, Wave 1 corresponds with a quickly increasing alarm function, reaching an almost 70\% reduction in transmission in six weeks in Miami and around 50\% reduction in transmission in Montr\'{e}al after eight weeks. Correspondingly, the reproductive number in both cities, which is estimated to be between 2 and 3 for both cities at the start of the pandemic decreases sharply, although it does not quite cross the threshold of one during the time period of modeling. Models which do not incorporate behavioral change through the alarm function estimate a constant reproductive number which is much lower (around 1.1 - 1.2), as these models do not contain the flexibility to allow transmission to change over time. During the second wave, the alarm functions no longer start at zero as it is not the start of the epidemic and cases and deaths are still occurring at low levels prior to the wave starting. In Miami, the alarm estimation during the second wave is heavily dependent on the inclusion of deaths in the alarm function, as our estimation of $\alpha$ showed deaths were the primary influence of behavioral change during this wave. Deaths were already elevated at the start of the wave, and alarm increases after around 40-50 days once the death rate started rapidly increasing in mid-July. The estimated reproductive number is about 1.25 in Miami and about 1.5 in Montr\'{e}al at the start of Wave 2. In both cities the behavioral change models do find that $\hat{\mathcal{R}_0}(t)$ becomes below 1 by the end of the modeling period. As in Wave 1, the models with no alarm results in a constant reproductive number at a much lower level of transmission.

\section{Discussion} \label{disc}

In this work we have extended Bayesian behavioral change epidemic models in three important ways to make them more realistic and accurate. We expanded the SIR compartmental structure to incorporate additional data sources on deaths and hospitalizations in an SIHRD models. To the best of our knowledge, this is the first instance of an expanded compartmental structure in a stochastic population-averaged model, where both the parameters associated with transition probabilities and all transitions themselves are given distributions and are imputed when not observed. We investigated the incorporation of undetected infections by assuming cases are observed according to a constant detection probability and the true number of infections is estimated using data-augmented MCMC. Finally, we extended the behavioral change framework of \cite{ward2023bayesian} to allow the population alarm to be influenced by cases and deaths. The proposed model specification also allows us to quantify the relative importance of each data source in informing population behavior, so it can be used to investigate potential changes in importance over time. 

Our  simulation study illustrated several important features of Bayesian behavioral change models. We showed that the SIHRD model offers better estimation of $\mathcal{R}_0$ and the relative importance parameter in the multivariable alarm, illustrating the benefit of expanding the compartmental structure so that all alarm inputs are endogenous. We also found that when deaths are highly important in influencing behavior through the population alarm, it is critical that deaths are incorporated into the alarm function, particularly for estimating the reproductive number towards the end of the epidemic. However, when deaths are equally as important as cases or minimally important, both the estimation of $\mathcal{R}_0$ and overall model fit are comparable between an alarm based on cases only or the more flexible multivariable alarm. 

Through analysis of multiple waves of COVID-19 in Miami and Montr\'{e}al, we illustrate the benefit of our modeling approach in practice. Ignoring behavioral change completely resulted in much poorer model fit than any of the behavioral change models we considered. We also found that cases were highly important in informing population behavior during the first wave of COVID-19 in both cities, but during the second wave in Miami, deaths were the primary influencing factor of behavior. This finding offers important insight into human behavioral change in response to a multiwave epidemic and may be due to perception going into Wave 2 that the virus was not as deadly as previously thought \citep{williams2021lessons}. The WAIC results from these analyses also indicated that there was minimal downside to using the more complex multivariable alarm function when only cases were important, and a large upside to including deaths when deaths were important. This is meaningful for using these models in practice as we do not know which data source is influencing behavior prior to model fitting.

Our data analysis also illustrated an important limitation of the current modeling framework. In the posterior predictive assessment of the SIHRD model fits for Miami during Wave 2, we found that the model adequately described the observed hospitalization and death trajectories, but the posterior predictive distribution of cases did not match well with the observed data. We hypothesize that this is due to assumption of exponentially distributed infectious and hospitalization periods for the SIHRD model, coupled with the use of strong priors which are needed for identifiability of the undetected infections as described in Section \ref{estimation}. More flexible distributions could be incorporated using the `linear chain trick' to convolve multiple exponential distributions \citep{cushing1997integrodifferential}, potentially in combination with strong priors informed by previously published literature, e.g. studies of the length-of-stay in the hospital for COVID-19 patients\citep{wen2022time}. Our implementation also assumes that the hospitalization rate is constant over the time period modeled, which may not be true as prior research has shown the COVID-19 hospitalization rate to vary over time \citep{couture2022estimating}. Our modeling framework would be amenable to varying the hospitalization rate over time, however, strong priors would still be required to incorporate undetected infections. Given that our primary interest was on describing and estimating behavioral change and that our data analysis focused on modeling relatively short time periods (six to ten weeks), we feel that the assumption of constant hospitalization rate over time is reasonable and the violation of this assumption may not be as impactful as it would be for modeling a longer duration.

In our application, we fit models to each wave separately. This approach reasonably accommodates changes in detection rates and transmissibility across time (e.g., from new variants), as these are likely to vary less within a relatively short time-frame, but could vary substantially when looking across an entire year or more of COVID-19 data. Modeling separate waves is also critical for inferring changes in the relative importance of cases or deaths across time, as the current modeling specification where the relative importance parameter, $\alpha$, is assumed to be constant. However, modeling multiple waves limits our ability to use the alarm function to explain long term behavior or capture carryover effects in behavior or transmission. Our behavioral change framework could be used for modeling multiple waves simultaneously, in which case we would want to extend the current model specification to allow for changes in the alarm specification, detection rate, and baseline transmission rate over time. Long-term behavioral change modeling may also necessitate the incorporation of behavioral fatigue, which is an interesting and ongoing area for future work.

Our SIR and SIHRD models have a few additional limitations that warrant discussion. First, they assume individuals are immediately infectious. While an Exposed compartment can be added to the behavioral change framework as in Ward et al\cite{ward2023bayesian}, this requires strong priors on the latent period and additional data augmentation. Because the SIHRD model already augments transitions to the removed compartments and undetected infections, we omitted the latent period to reduce computational complexity. Second, we do not allow transmission from hospitalized individuals. Although a hospital transmission term could be added to the transmission probability, such effects are typically small, likely unidentifiable from available data, and would complicate interpretation of the alarm function. Third, deaths are a lagged consequence of cases, which may limit the ability of our alarm function to fully disentangle responses due to deaths from delayed responses due to cases. We believe that using 30-day smoothed case and death counts in our data analyses mitigates this issue to a large extent as it incorporates a lag in behavioral change in response to both cases and deaths. However, these effects would be challenging to fully disentangle and may require explicit modeling of the lag on each data source or exploring alternative lag specifications. Finally, we assume a constant detection rate within the modeled time period, while in reality studies indicate that COVID-19 case ascertainment varied over time \citep{russell2020reconstructing, irons2021estimating}. One way to address this is to model multiple waves of relatively short time periods separately, as we did in our data analysis. Incorporating changes in detection within a wave is more challenging when we are simultaneously estimating changes in transmission over time due to behavioral changes, due to the partial identifiability of these models. To overcome this challenge, one could try to incorporate auxiliary data sources such as wastewater \citep{mcmahan2021covid} or seroprevalence surveys \citep{irons2021estimating}, and this is an avenue for future work for Bayesian behavioral change models.

Our specified formulation of the SIHRD model allows both undetected infections and detected cases to become hospitalized. This requires strong priors on the parameters governing transitions out of the infectious compartment to induce identifiability as described in Section \ref{estimation}. Alternatively, in some cases it may be reasonable to assume that hospitalizations only arise from detected cases. We have detailed this alternative formulation and investigated some important properties in simulations in the Supplementary Material Section 4. We found that the modified model does improve identifiability of the parameters $\lambda$ (I to H) and $\gamma_1$ (I to R), and that estimation of the reproductive number or alarm function parameters was not impacted, suggesting that this model is reasonable alternative to our SIHRD implementation. However, the assumption that hospitalizations to only arise from observed cases is difficult to assess in practice as publicly available data provide only counts of the numbers of new cases and hospitalizations on each day, without a link between the specific individuals appearing in each set of counts.

\section*{Software}

Software in the form of R code, together with data used and complete documentation is available at \url{https://github.com/ceward18/multipleDataBCM}.

\section*{Acknowledgments}

Funding for the project was provided by the Canadian Statistical Sciences Institute (CANSSI) Distinguished Postdoctoral Fellowship. A. M. Schmidt and R. Deardon acknowledge financial support from the Natural Sciences and Engineering Research Council (NSERC) of Canada Discovery Grants program (Schmidt—RGPIN/2024-04312, Deardon—RGPIN/03292-2022). The authors also acknowledge the Minnesota Supercomputing Institute (MSI) at the University of Minnesota for providing resources that contributed to the research results reported within this paper. URL: \url{http://www.msi.umn.edu}.

\vfill

\pagebreak

\bibliography{references}

\end{document}